\documentstyle[prd,aps,floats,twocolumn]{revtex}
\def\be{\begin{equation}}
\def\ee{\end{equation}}
\def\bea{\begin{eqnarray}}
\def\eea{\end{eqnarray}}

\newcommand\lsim{\mathrel{\rlap{\lower4pt\hbox{\hskip1pt$\sim$}}
    \raise1pt\hbox{$<$}}}
\newcommand\gsim{\mathrel{\rlap{\lower4pt\hbox{\hskip1pt$\sim$}}
    \raise1pt\hbox{$>$}}}

\begin{document}
\preprint{} 
\draft

%
%
\input epsf
\renewcommand{\topfraction}{0.99}
\renewcommand{\bottomfraction}{0.99}
\twocolumn[\hsize\textwidth\columnwidth\hsize\csname@twocolumnfalse\endcsname

\title{On the Amount of Gravitational Waves from Inflation}
\author{L. Pilo$^1$, A. Riotto$^1$ and A. Zaffaroni$^2$}
\address{(1) Department of Physics and INFN
Sezione di Padova, via Marzolo 8, I-35131 Padova, Italy}
\address{(2) Universit\'a di Milano-Bicocca and INFN, Piazza della Scienza 3, 
Milano I-20126, Italy}

\date{\today}
\maketitle
\begin{abstract}
\noindent
The curvaton and the inhomogeneous reheating scenarios 
for the generation of the cosmological
curvature perturbation on large scales represent an alternative to the standard
slow-roll scenario  where the observed density 
perturbations are due to fluctuations of the inflaton field itself. The 
basic assumption  of the curvaton and inhomogeneous reheating 
mechanisms is that the initial
curvature perturbation due to the inflaton field is negligible. 
This is usually attained by lowering the energy scale of inflation 
thereby concluding that  the amount of gravitational waves produced during
inflation is highly suppressed. We show, contrary to this common lore, that
the curvaton and the  inhomogeneous reheating scenarios are compatible
with a level of  gravity-wave fluctuations which may well be
observed in future satellite experiments. 
This conclusion does not involve, as in the slow-roll inflationary models, 
embarrasingly large
field variations in units of the Planck scale.
As a working example,
we illustrate the recently proposed stringy version of old inflation.

\end{abstract}


\vskip2pc]


\noindent
Inflation \cite{guth81,lrreview} has  become the dominant 
paradigm for understanding the 
initial conditions for structure formation and for Cosmic
Microwave Background (CMB) anisotropy. In the
inflationary picture, primordial density and gravity-wave fluctuations are
created from quantum fluctuations ``redshifted'' out of the horizon during an
early period of superluminal expansion of the universe, where they
are ``frozen'' \cite{muk81,bardeen83}. 
Perturbations at the surface of last scattering are observable as temperature 
anisotropy in the CMB, which was first detected by the Cosmic Background 
Explorer (COBE) satellite \cite{bennett96,gorski96}.
The last and most impressive confirmation of the inflationary paradigm has 
been recently provided by the data 
of the Wilkinson Microwave Anistropy Probe (WMAP) mission which has 
marked the beginning of the precision era of the CMB measurements in space
\cite{wmap1}.
The WMAP collaboration has  produced a full-sky map of the angular variations 
of the CMB, with unprecedented accuracy.
WMAP data confirm the inflationary mechanism as responsible for the
generation of curvature (adiabatic) super-horizon fluctuations. 

Despite the simplicity of the inflationary paradigm, the mechanism
by which  cosmological adiabatic perturbations are generated  is not
yet established. In the standard slow-roll scenario associated
to one-single field models of inflation, the observed density 
perturbations are due to fluctuations of the inflaton field itself when it
slowly rolls down along its potential. 
When inflation ends, the inflaton $\phi$ oscillates about the minimum of its
potential $V(\phi)$  and decays, thereby reheating the universe. As a result of the 
fluctuations
each region of the universe goes through the same history but at slightly
different times. The 
final temperature anisotropies are caused by the fact that
inflation lasts different amounts of time in different regions of the universe
leading to adiabatic perturbations. Under this hypothesis, 
the WMAP dataset already allows
to extract the parameters relevant 
for distinguishing among single-field inflation models \cite{ex}.

An alternative to the standard scenario is represented by the curvaton 
mechanism
\cite{curvaton1,LW,curvaton3,LUW} where the final curvature perturbations
are produced from an initial isocurvature perturbation associated to the
quantum fluctuations of a light scalar field (other than the inflaton), 
the curvaton, whose energy density is negligible during inflation. The 
curvaton isocurvature perturbations are transformed into adiabatic
ones when the curvaton decays into radiation much after the end 
of inflation. 
Recently, another mechanism for the generation of cosmological
perturbations has been proposed \cite{gamma1,gamma2,gamma3}
dubbed the inhomogeneous reheating scenario.  
It acts during the reheating
stage after inflation if super-horizon spatial
fluctuations in the decay rate of the inflaton field
are induced during inflation, causing  adiabatic perturbations
in  the final reheating temperature
in different regions of the universe.

Contrary to the standard picture, the curvaton and the 
inhomogeneous reheating 
mechanisms exploit the fact that 
the total curvature perturbation (on uniform density hypersurfaces)
$\zeta$ can change on arbitrarily large scales due to a non-adiabatic
pressure perturbation    which may be 
present  in a multi-fluid system \cite{Mollerach,MFB,GBW,WMLL,wmu}.
While the entropy
perturbations evolve independently of the curvature perturbation on
large scales,  the evolution of the large-scale curvature is
sourced by entropy perturbations. 

The generation of gravity-wave fluctuations 
is a generic prediction of an accelerated  de Sitter expansion of the universe.
Gravitational waves, whose possible observation might come from the  
detection of  the $B$-mode of polarization in the
CMB anisotropy \cite{polreview},   
may be viewed as  ripples of spacetime around the  background metric
$g_{\mu\nu}=dt^2-a^2(t)\left(\delta_{ij}+h_{ij}\right)
dx^i dx^j$, where $a$ is the scale factor and $t$ is the cosmic time. The  
tensor $h_{ij}$ is traceless and transverse and has two degrees polarizations,
$\lambda=\pm$.
Since gravity-wave fluctuations are (nearly)
frozen on superhorizon scales,
a way of characterizing them is to compute
their spectrum on scales larger than the horizon. 

In the standard slow-roll inflationary models where the fluctuations 
of the inflaton field
$\phi$ are responsible for the curvature perturbations, the 
power spectrum of gravity-wave  modes is 

\begin{equation}
{\cal P}_{T}(k)=\frac{k^3}{2\pi^2}\sum_{\lambda}\left|
h_{\bf k}
\right|^2=\frac{8}{M_p^2}\left(\frac{H_*}{2\pi}\right)^2
\left(\frac{k}{aH_*}\right)^{-2\epsilon}\, ,
\end{equation}
where $M_p=(8\pi G)^{-1/2}\simeq 2.4\times 10^{18}$ GeV 
is the Planck scale. Here
$\epsilon=(\dot{\phi}^2/2M_p^2 H_*^2)$
is a standard  slow-roll parameter and $H_*=\dot a/a$
indicates the Hubble rate during inflation. On the other hand, 
the power spectrum of curvature
perturbations in slow-roll inflationary models is given by

\begin{equation}
{\cal P}_{\zeta}(k)=
\frac{1}{2 M_p^2\epsilon}\left(\frac{H_*}{2\pi}\right)^2
\left(\frac{k}{aH_*}\right)^{n_\zeta-1}\, ,
\end{equation}
where $n_{\zeta}\simeq 1$ is the spectral index. Since the fractional changes
of the power spectra with  scales are  much smaller than unity, one can safely 
consider the power spectra as roughly constant on the  scales relevant for the
CMB anisotropy and define a tensor-to-scalar amplitude ratio 

\begin{equation}
\label{q}
r=\frac{{\cal P}_{T}}{16\,{\cal P}_{\zeta}}=\epsilon\, .
\end{equation}
The present WMAP dataset  allows
to extract the upper bound $r\lsim 10^{-1}$ \cite{ex}. 
Since  the scale of inflation 
in slow-roll models of inflation is fixed to be

\begin{equation}
\label{cobe}
V^{1/4}\simeq 6.7\,r^{1/4}\times 10^{16}\,{\rm GeV}\, ,
\end{equation}
in order to match the observed amplitude of CMB anisotropy, 
one can already infer an upper bound on the energy scale of inflation
of about $4\times 10^{16}\,{\rm GeV}$. 
The corresponding upper bound on the Hubble rate
during inflation $H_*$ is about $4\times 10^{14}$ GeV.
What is more relevant though is the width
of the potential \cite{lythgrav}. The slow-roll paradigm gives, using
the definition of $\epsilon$ and 
Eq. (\ref{q}), 

\begin{equation}
\label{k}
\frac{1}{M_p}\left|\frac{d\phi}{dN}\right|=\sqrt{2}\,r^{1/2}\, ,
\end{equation}
where $d\phi$ is the change in the inflaton field in $dN=Hdt\simeq
d\,{\rm ln}\,a$ Hubble times. While the scales corresponding to
the relevant multipoles in the CMB anisotropy are living the horizon
$\Delta N\simeq 4.6$ and therefore the field variation
is $
(\Delta\phi/M_p)\simeq \left(r/2\times 10^{-2}\right)^{1/2}$
\cite{gravex}. This is a minimum
estimate because inflation continues for some number $N$ of $e$-folds
of order of 50.
The detection of   gravitational waves requires in general
variation of the inflaton field of the order of the Planck scale \cite{lythgrav}. 
This conclusion
sounds fairly pessimistic since slow-roll models of inflation are generically
based on four-dimensional field theories, possibly involving supergravity,
where higher-order operators with powers of $(\phi/M_p)$ are disregarded. This
assumption is justified only if the inflaton variation is small
compared to the Planck scale. It is therefore difficult to 
construct a satisfactory model of inflation firmly rooted  
in modern particle theories having supersymmetry as a
crucial ingredient and with large variation of the inflaton field.
There is  a strong theoretical
prejudice against the likelyhood of observation of gravity-wave
detection within slow-roll models of inflation where the
curvature perturbation is due to the fluctuations of the
inflaton field itself.

What about the expected amplitude of gravity-wave fluctuations in 
the other scenarios where the curvature perturbation is generated through
the quantum fluctuations of a scalar field other than the inflaton? Consider,
for instance, the curvaton scenario \cite{LW}. 
During inflation, the curvaton energy density is 
negligible and isocurvature perturbations with
a flat spectrum are produced in the curvaton field $\sigma$,
$\langle\delta\sigma^2\rangle^{\frac{1}{2}} = (H_*/2\pi)$, 
where $\sigma_*$ is the value of the curvaton field during inflation.
After the end
of inflation, 
the curvaton field oscillates during some radiation-dominated era,
causing its energy density to grow and 
thereby converting the initial isocurvature into curvature 
perturbation. Indeed, after a few Hubble times the curvaton 
oscillation will be sinusoidal
except for the Hubble damping. The  energy density $\rho_\sigma$
will then be proportional to the square of the oscillation amplitude,
and will scale like the inverse of the locally-defined comoving volume
corresponding to matter domination. On the spatially flat slicing, 
corresponding to uniform local expansion, its perturbation has a constant
value $
\delta\rho_\sigma/\rho_\sigma \simeq \left(\delta\sigma/\sigma 
\right)_*$.
The curvature perturbation $\zeta$ is supposed to be negligible when the
curvaton starts to oscillate, growing during some radiation-dominated
era when $\rho_\sigma/\rho\propto a$.
After the curvaton decays $\zeta$  becomes constant. In 
the approximation that the curvaton decays instantly
it is then given by $
\zeta \simeq (2\gamma/3) \left(\delta \sigma/\sigma \right)_*$, 
where $\gamma\equiv (\rho_\sigma/\rho)_{D}$ 
and the subscript $D$ denotes the epoch of decay. The corresponding spectrum
is \cite{LW}

\begin{equation}
\label{spectrum}
{\cal P}_\zeta^{\frac{1}{2}}\simeq
\frac{2\gamma}{3}  \left(\frac{H_*}{2\pi \sigma_*}\right)\, .
\end{equation}
The curvaton  scenario -- as well as the inhomogeneous reheating scenario --
 liberates the inflaton from the responsibility
of generating the cosmological curvature perturbation
and therefore  avoids slow-roll conditions. Its
 basic assumption  is that the initial
curvature perturbation due to the inflaton field is negligible. 

The common lore is to assume that the energy scale of the inflaton potential
is too small to match the observed amplitude of CMB anisotropy, 
that is

\begin{equation}
\label{curvaton}
V^{1/4}\ll  6.7\,r^{1/4}\times 10^{16}\,{\rm GeV}\, ,
\end{equation}
corresponding to $H_*\ll 10^{14}$ GeV. Therefore --  while
certainly useful to construct low scale models
of inflation \cite{lythdim} -- 
it is usually thought that the curvaton
mechanism (as well as the inhomogeneous reheating scenario) predicts
an amplitude  of gravitational waves which is far too small to be
detectable by future satellite experiments aimed to
observe the $B$-mode of the CMB polarization. 

This conclusion would be
discouraging if true.  It would imply that all  future efforts in
measuring tensor modes in the CMB anisotropy are of no use
because  both the slow-roll standard scenario (for theoretical
reasons) and the alternative
mechanisms for the production of the curvature perturbation
predict a tiny level of gravity-wave fluctuations. 

There is however an obvious  way out to  such a pessimistic
scenario.   Within the curvaton mechanism (and the 
inhomogeneous reheating scenario), the curvature perturbations
of the inflaton field may be highly suppressed not because the energy
scale of is tiny, but 
because
the inflaton field  is well anchored at the false vacuum 
with a mass much $m_\phi$ larger than the Hubble rate during inflation.
Suppose that the inflaton potential is of the form

\begin{equation}
V(\phi)= V_0+\frac{m_\phi^2}{2}(\phi-\phi_0)^2+\cdots\, ,
\end{equation}
where  $\phi_0$ is the location of the minimum around which
$m_\phi\gg H_*$.
Under these circumstances, slow-roll conditions are badly violated
since $\eta=(m_\phi^2/3 H_*^2)\gg 1$
and 
the fluctuations of the inflaton 
field on super-horizon scales read \cite{tonireview}

\begin{eqnarray}
\left|{\delta \phi}_{\bf k}\right|^2&\simeq&\frac{\pi}{4}
\frac{e^{-\pi\nu}}{a^2} \frac{1}{aH}\left[
\frac{(1+{\rm coth}(\pi\nu))^2{\rm sinh}(\pi\nu)}
{\pi\nu}\right.\nonumber\\
&+&\left. \frac{\nu}{\pi
{\rm sinh}(\pi\nu)}\right]\, ,
\end{eqnarray}
where $\nu=m_\phi/H_*$. The resulting power spectrum, after properly substructing
the zero-point fluctuations, is highly suppressed 

\begin{equation}
{\cal P}_{\delta \phi}(k)=
\Big(\frac{H_*}{2\pi}\Big)^2 \Big(\frac{k}{aH_*}\Big)^3\, e^{-2 m_\phi^2/H_*^2}\, .
\end{equation}
Thus, the embarassing constraint (\ref{curvaton}) is circumvented
 and  the curvature perturbations due to the inflaton field may be
small even if the Hubble
rate during inflation is  sizeable (even close
to the present upper bound of  $\sim 4\times 10^{14}$ GeV). 

We conclude
that within the curvaton scenario (and the inhomogeneous reheating scenario)
gravity-wave fluctuations may well be
observed in future satellite experiments. This conclusion does not imply 
embarassingly large field
variations in units of the Planck scale as in the slow-roll
scenario.
A
future detection of the $B$-mode
of CMB polarization will not automatically disprove the 
curvaton and the homogeneous reheating scenarios.
This is our main
observation. 

Let us elaborate on this point.
Since the amplitude of curvature perturbation ${\cal P}_\zeta^{1/2}$ 
must
match the observed  value $5\times
10^{-5}$, from Eq. (\ref{spectrum}) one infers that

\begin{equation}
\sigma_*\simeq 2\,\gamma\,\times 10^3\, H_*\, .
\end{equation}
On the other hand, a large signal-to-noise ratio  for the
detection of the $B$-mode of CMB 
polarization requires $H_*\gg 2\times 10^{12}$ GeV
\cite{gravex} (notice that for $H_*\lsim 10^{11}$ GeV 
secondary effect dominates over that of primordial tensors \cite{lll})

\begin{equation}
\label{p}
\sigma_*\gg   20\, H_*\gsim 10^{-5}\, M_p\, ,
\end{equation}
where in the last inequality we have made use of the 
current
WMAP bound on non-gaussianity      which 
requires $\gamma\gsim 9\times 10^{-3}$ \cite{ng}. If future WMAP data
on non-gaussianity  strengthen the bound on $\gamma$, the lower bound
on $\sigma_*$ will be corrispondingly tighter.
Similarly, the present
upper bound on the Hubble rate, $H_*\lsim 4\times 10^{14}$ GeV \cite{ex}, implies

\begin{equation}
\label{pp}
\sigma_*\lsim   2\times 10^3\, H_*\lsim  0.3\, M_p\, ,
\end{equation}
where we have taken into account that $\gamma$ has to be smaller than unity.

Curvaton scenarios able to predict a detectable background of gravity-wave
fluctuations are characterized by large values of the Hubble rate and $\sigma_*$, 
$10^{-5}\ll \sigma_*/M_p\lsim  0.3$.
A working example of a model of inflation 
manifesting these properties is 
the  recently proposed   
stringy version of  old inflation  \cite{noi}. 
The model does not require any slow-roll inflaton potential and  is 
based on string compatifications with warped metric. 
Warped factors are quite common in string theory compactifications
and arise, for example, in the vicinity of $D$-branes sources.
Similarly, string theory has antisymmetric forms whose fluxes in the 
internal
directions of the compactification typically introduce warping.
From the inflationary point of view, the basic property of the set-up
is that it admits in the deep IR region
of the metric the presence of $p$ anti-$D$3-branes. These anti-branes
generate a positive vacuum energy density. For sufficiently
small values of $p$, the system sits indeed on a false vacuum state
with positive  vacuum energy density. The latter is responsible 
for the accelerated period of inflation. 

In terms of the four-dimensional
effective description, the inflaton may be identified with
a four-dimensional scalar field parameterizing the angular
position $\phi$ of the anti-$D$3-branes in the internal directions.
The mass squared of the inflaton at the  false
ground state is $m_\phi^2\sim (1/\alpha^\prime )(r_0/R)^2$, where
$\alpha'$ is the string length squared and 
$(r_0/R)$ is the warp factor ($R$ is the AdS radius and $r_0$
is the location of the stack on antibranes along the
fifth dimension). The mass squard is much larger than 

\begin{equation} 
H_*^2\simeq 2\,p\,\frac{T_3}{3M_p^2}\left(\frac{r_0}{R}\right)^4\, ,
\end{equation}
where $T_3$ is the tension of a single   anti-$D$3-brane. The false vacuum
for the potential $V(\phi)$ 
exists only if the number of anti-$D$3-branes is smaller than
a critical number \cite{KV}. If a sufficient number of anti-$D$3-branes
travels from the UV  towards the IR region,
thus increasing the value of $p$, inflation
stops as soon as $p$ becomes larger  than the critical value $p_{cr}$. 
Once an  anti-$D$3-brane appears in the UV region, it rapidly flows towards
the IR region 
and it starts
oscillating with a  frequency $\sim \sqrt{2} H_*$ 
around the minimum where the $p$ anti-$D$3-branes
sit. 
 Since the universe is 
in a de Sitter phase, the amplitude of the oscillations decreases as rapidly 
as $\sim e^{-3N/2}$. Once the energy density stored in the
oscillations becomes smaller than $\sim 1/\alpha^{'2}$ the anti-$D$3-brane
stops its motion and gets glued with the $p$ anti-$D$3-branes in the IR increasing
thier number by one unity. 
The oscillating phase lasts for a number of $e$-folds of the order of
$\sim{2\over 3}{\rm ln}
(p^3\left(r_0/R\right)^4/ M_p^2\alpha^\prime)$ \cite{noi}.
Once
the number of anti-$D$3-brane becomes equal to $p_{cr}$, 
inflation ends since 
the curvature around the false vacuum becomes negative.  The system
rolls down the supersymmetric vacuum and the graceful exit from inflation 
is attained. In such a 
stringy version of the old inflation it is possible
to find scalar fields which have all the necessary properties to
play the role of the curvaton. For instance, the imaginary part
of the volume modulus behaves like an axion field and its mass
is exponentially suppressed upon volume stabilization. 
Because of this exponential suppression, 
the condition $m_\sigma^2\ll H_*^2$ during inflation does not
require any particular fine-tuning \cite{noi}.
Furthermore, since the non-perturbative superpotentials necessary to
volume stabilization 
arise in the UV region, no warping suppression is expected and the
curvaton  scale $\sigma_*$ will be naturally of the order of $M_p$ 
in the four-dimensional effective theory.
Furthermore, since $p$ has to be larger than about $(R/r_0)^4$ in order for the
$p$ anti-$D$3-branes to dominate the energy density of the universe, the
Hubble rate during inflation is naturally large

\begin{equation}
H^2_*\gsim \frac{T_3}{M_p^2}\sim 
10^{-3}\frac{1}{g_s\alpha'^2 M_p^2}\, ,
\end{equation}
where $g_s$ is the string coupling constant.
For $g_s\sim 0.1$, requiring that $H_*$ is larger than
$\sim 2\times 10^{12}$ GeV to ensure gravity-wave detection
\cite{gravex} implies the fundamental scale $1/\sqrt{\alpha'}$ to be of the
order of $10^{16}$ GeV, a perfectly natural value.


\end{document}